# Solving equations after dense scan to improve the resolutions of microscopes

Yaohua Xie, Yaohua.Xie@hotmail.com
(ORCID: 0000-0001-6780-3156)

**Abstract** Super-resolution techniques overcome the diffraction-limit and get very high resolutions. A category of these techniques, e.g., STED achieves this by creating an illumination spot smaller than the Airy Disk. As a result, points are distinguishable even if they are as small as the spot. In order to further observe structures smaller than the spot itself, a technique called DDS scans the sample more densely, and recovers the expected image by deconvolution. In that technique, the deconvolution is achieved by filtering which requires some peripheral areas to be scanned together with the region of interest. In this study, an approach is proposed which has the same preprocessing stage as DDS. But it requires to scan only the region of interest. After that, an equation system is got from the scanned data. Finally, the expected image is recovered by solving the equation system. Experiments are performed on simulated data, and the results demonstrate the effectiveness of the proposed approach. The experiments also suggest that the proposed approach is more efficient than the existing one especially when the expected resolution is high.

The diffraction-limit was once a great barrier in the development of light microscope. This barrier was not broken until the technique of super-resolution emerged after more than one hundred years(Sigal, Zhou et al. 2018). Some super-resolution techniques, such as STED (stimulated emission depletion) microscope(Hell and Wichmann 1994), use very tiny spot to illuminate (scan) samples. The spot is smaller than the Airy-disk, i.e., the central area of the PSF (Point Spread Function) of light microscope. By scanning the sample spot by spot, the sample's structure can be resolved even if it is much smaller than the diffraction-limit. In order to further resolve structures smaller than the spot itself, a technique called DDS (Deconvolution after Dense Scan) has been proposed recently(Xie 2019). It first determines the ROI (Region of Interest) of the sample, and eliminates the optical uncertainty of the ROI's peripheral areas. Then, the sample is scanned with denser step than in conventional approaches. Finally, sharp and high resolution images are got by deconvolution. In that technique, the deconvolution is achieved by filtering, such as inverse filtering or wiener filtering(Gonzalez and Woods 2014). These are effective ways of deconvolution, but they require the peripheral areas to be scanned. That will cause extra work especially when the peripheral areas are large.

Therefore, we propose to perform deconvolution by solving equation systems instead of filtering. By doing so, only the ROI needs to be scanned, and then an equation system can be built on the scanned data. The solution of the equation system is exactly the sharp and high resolution image. The proposed technique is termed "Solving Equations after Dense Scan (SEDS)".

In total, the proposed approach can also be divided into three major stages. The first stage is basically the same as that in DDS. In this stage, the sample is pre-processed to eliminate the optical uncertainty of the peripheral areas around the ROI. As a result, the optical property of those areas is known after this stage. In different applications, optical property may mean light reflection or fluorescence excitation property. Optical uncertainty could be eliminated in various ways. For example, the light reflection and fluorescence excitation is made zero (or ignorable) in the peripheral areas. This pre-processing stage provides the pre-condition of the following stages. Fig. 1 shows the spot,

ROI and peripheral areas. Where, the bright circle represents the spot, the textured rectangle represents the ROI, and the dark border represents its peripheral areas. In this technique, the peripheral areas should be at least as wide as the spot's radius, or half of its size.

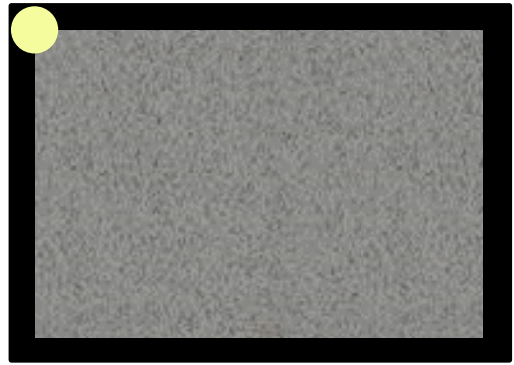

Fig.1. The spot, ROI and peripheral areas

The second stage is the dense scan of the sample. In DDS, the sample is scanned densely by the illumination spot. Each time, a tiny part of the sample is illuminated by the spot, and its image is acquired. Then, a value is calculated from the image, which represents the optical property of the illuminated part. After the ROI and its peripheral areas are all scanned, all such values are combined to generate an intermediate image. Dense scan means the scanning step is smaller than the spot's size, thereby the intermediate image appears blurred. But it implicitly includes the detail information of the sample's structure, and thereby the expected image can be recovered by deconvolution. In this study, the sample is also scanned densely and each value is calculated according. But only the ROI is scanned, and not the peripheral areas. Then, the calculated values are not combined to get an intermediate image, but are used to build an equation system instead. The equation system also implicitly includes the sample's detail information, and the expected image can be recovered by solving the equation system. Fig. 2 shows the "footprints" of the spot in DDS and this study, respectively. In each figure, the dash-line rectangle represent the ROI area, and the spot's footprints partly overlap one another. It can be seen that the spot's center only lies within the ROI in this study. This will save a lot of work when the spot is much larger than the ROI (or, when the expected resolution is very high).

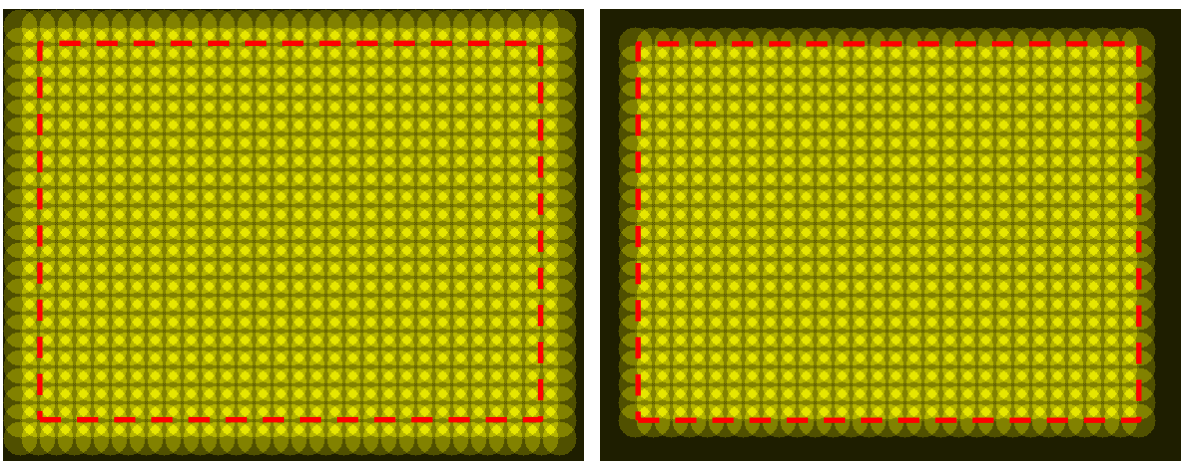

Fig.2. (a) Footprints in DDS　　　　　　　　　　(b) Footprints in this study

In the third stage of DDS, the expected image is recovered from the intermediate image. Since the intermediate image is equivalent to the convolution of the expected image with the spot, the recovery is done by deconvolution. In this study, an equation system is built during the second stage instead of an intermediate image. It is comprised of the pixels of the expected image (ROI). When part of the sample is illuminated by the spot, each point in the resulting image is the product of the point's optical property with the corresponding illumination. In computer, the resulting image can be modeled as a "product image", which is the product of the expected image's pixel with the corresponding spot image's pixel. After the resulting image is formed by the microscope, the sum of its light intensity is calculated. Corresponding to the above computational model, the sum of the product image's pixels is calculated. Such a relationship can be represented by the following equation system:

$$S(i,j) = \sum_u \sum_v I(u,v) \cdot E(i+u, j+v) \tag{1}$$

Equation (1) shows the relationship about the product at row $i$ and column $j$. Assume that the illumination spot is centered at the pixel $(i,j)$ of the expected image. The illumination spot is denoted as $I$, and the expected image is denoted as $E$. When $I$ multiplies $E$ pixel-wise, the sum of the resulting product is denoted as $S(i,j)$. If the expected image has $R$ rows and $C$ columns, equation system (1) can be written as the following detailed form:

$$\begin{cases} S(1,1) = \sum_u \sum_v I(u,v) \cdot E(1+u, 1+v) \\ \vdots \\ S(1,C) = \sum_u \sum_v I(u,v) \cdot E(1+u, C+v) \\ \vdots \\ S(i,j) = \sum_u \sum_v I(u,v) \cdot E(i+u, j+v) \\ \vdots \\ S(R,1) = \sum_u \sum_v I(u,v) \cdot E(R+u, 1+v) \\ \vdots \\ S(R,C) = \sum_u \sum_v I(u,v) \cdot E(R+u, C+v) \end{cases} \tag{2}$$

There are $R \cdot C$ unknowns in the above equation system, i.e., $E(i,j)$, where $i = 1,2,\ldots,R$ and $j = 1,2,\ldots,C$. There are also $R \cdot C$ equations in it. Thereby, it is possible to solve the unknowns(Lay, Lay et al. 2015). If the peripheral areas are also scanned and more equations are built, that may be beneficial to solve it(Sheffield 2019). Assume that the optical property of the ROI's peripheral areas is made zero in the pre-processing stage. Thereby, $E(i+u, j+v)$ is treated as zero when $i+u$ is beyond $[1,R]$ and/or $j+v$ is beyond $[1,C]$. If the optical property is other values instead of zero, $E(i+u, j+v)$ should be set the corresponding values.

When the equation system is being solved by computer, usually it should be written in matrix form: $Ax = b$. Where, $A$ is comprised of $I(u,v)$, $x$ is comprised of $E(i+u, j+v)$, and $b$ is comprised of $S(i,j)$. In other words, $x$ is comprised of the expected image's pixels. Fig. 3(a) and 3(b) visualizes an example value of matrix $A$ and vector $b$, respectively. The data come from an experiment simulating the resolution of 0.1 nm/pixel. These figures looks meaningless and almost no information about the sample's structure can be seen directly. But they are actually informative enough, and includes full information about the sample's structure.

The above procedure gives a generic solution, but it may not be suitable for some extreme cases. For example, the equation system is not solvable if the spot has a constant value and is much larger than the ROI. In this case, the spot's light intensity is the same at any point, thereby all the $I(u,v)$ in the above equations are the same. As a result, all the $R \cdot C$ equations are actually the same, and cannot be solved uniquely. Fortunately, such an extreme case is rare in real world. Actually, it could be difficult to "create" such an illumination spot even if intentionally.

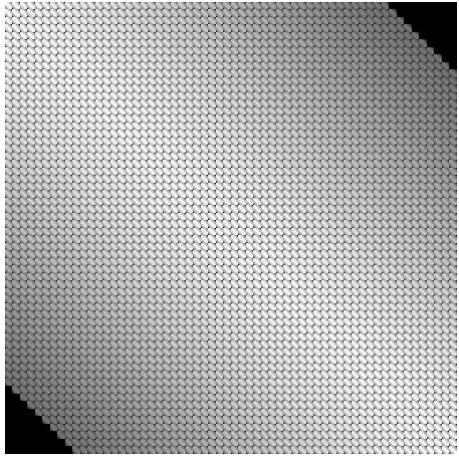 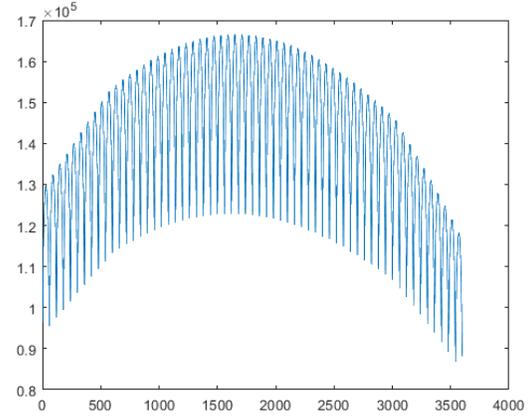

Fig.3. (a) Example value of matrix $A$　　　　　　(b) Example value of vector $b$

Two simulation experiments have been performed to test the effectiveness of the proposed technique. In the first experiment, the size (diameter) of the illumination spot is assumed to be 9 nm, and the scanning step is set to 3 nm. Thereby, the expected resolution is also 3 nm/pixel in this experiment. As a result, the illumination spot is 3x3 pixels, and the Airy disk's radius is between [67, 100] pixels in this case. The expected image is the simulation of the physical sample, and its size equals the total number of scanned footprint. It is used in a simulated imaging procedure, and also used for verifying the accuracy of recovery. Fig.4 shows the key data of the simulation experiment. Where, Fig. 4(a) is the expected image, Fig. 4(b) is the result of conventional microscope, Fig. 4(c) is the result of STED, and Fig. 4(d) is the image recovered by the proposed technique. It can be seen that the conventional result is very blurred, and basically no details can be observed directly. The result of STED is as sharp as the expected image, but its size is only 1/3 of the latter because the spot is 3x3 pixels. The image recovered by the proposed technique is almost the same as the expected image. Actually, the averaged difference of pixels between the two images is about 8.03e-12.

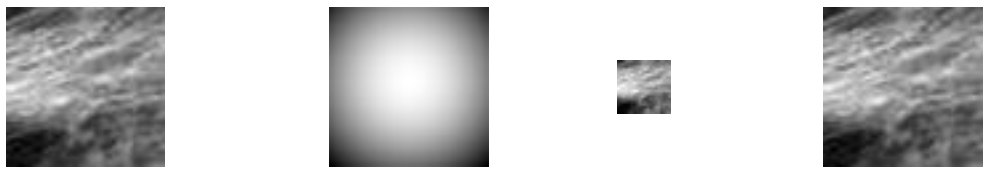

Fig.4. (a) Expected image　　(b) Conventional result　　(c) STED result　　(d) Recovered image

In the second experiment, the size (diameter) of the illumination spot is assumed to be 10.1 nm, and the scanning step is set to 0.1 nm. Thereby, the expected resolution is also 0.1 nm/pixel in this experiment. As a result, the illumination spot is 101x101 pixels, and the Airy disk's radius is between [2000, 3000] pixels in this case. Fig.5 shows the key data of this simulation experiment. Where, Fig. 5(a) is the expected image, Fig. 5(b) is the result of conventional microscope, Fig. 5(c) is the illumination spot, and Fig. 5(d) is the image recovered by the proposed technique. In this experiment, the conventional result is even more blurred. Besides, the ROI is smaller than the spot, thereby no image is got by STED. Similar to the first experiment, the image recovered by the proposed technique is almost the same as the expected image. The averaged difference of pixels is about 1.11e-06, which is about 0.0000016% of the mean pixel value of the expected image. This difference is larger than the first experiment, but still very small and ignorable.

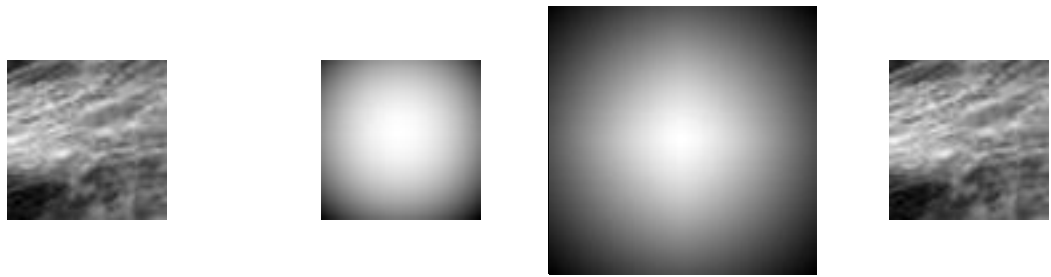

Fig.5. (a) Expected image    (b) Conventional result    (c) Illumination spot    (d) Recovered image

Compared to DDS, the proposed approach does not scan the peripheral areas. For the first experiment, the expected image is 60x60 pixels and the spot is 3x3 pixels. Thereby, the total number of scanned footprint is 64x64 when using DDS. But only 60x60 footprints need to be scanned in the proposed approach. For the second experiment, the expected image is still 60x60 pixels but the spot is 101x101 pixels. Thereby, the total number of scanned footprint is 260x260 when using DDS. But still only 60x60 footprints are enough in the proposed approach. Usually, the larger the spot, the more footprints need to be scanned in DDS. But in the proposed approach, the number of necessary footprint is only the same as the expected image's pixel number. That suggests the proposed technique might be more efficient when the expected resolution is higher. Limited by computational resources, experiments are not performed on higher resolutions. But the proposed approach can actually achieve unlimited high resolution in principle.

In summary, an existing technique, DDS is modified in this study. Only the ROI area is scanned densely in the proposed approach, and not the peripheral areas. Then, an equation system is built from the scanned data, and finally the expected image is recovered. According to its principle, the proposed approach would be more efficient especially when the expected resolution is high. Two simulation experiments have been performed, and the results demonstrate the effectiveness of the proposed approach.